\documentclass[12pt]{article}
\usepackage{a4}
\usepackage{epsfig}
\begin{document}
\title{Surface Phason--Polaritons
in Charge Density Wave Films}

\author {W. Wonneberger and R. Lamche\\
\\
{Department of Physics, University of Ulm}\\
{D--89069 Ulm, Germany\/}}

\date{}
\maketitle

\begin{abstract}
\noindent
The coupled non--radiative excitations of the electromagnetic field
and phasons in films with a quasi one--dimensional
charge density wave (CDW) are evaluated for P--polarization and
CDW conducting axis inside the film. The prominent features are two
surface phason--polariton branches extending from the CDW pinning
frequency to the frequency of the longitudinal optical phason. These
surface phason--polariton states are confined to a finite band of
longitudinal wave numbers. Besides surface polaritons, infinite
series of guided wave modes are found which extend to large wave
numbers. These differences to usual phonon--polaritons are caused
by the extreme anisotropy of the electric CDW reponse. At finite
temperatures, quasi--particles are thermally excited. Their
dissipation leads to polariton damping. Significant level shifts
of surface phason--polaritons and guided wave modes are also found.
They are due to thermal dressing of the longitudinal optical phason
via quasi--particles. The zero and finite temperature
results including the case of neglected retardation are displayed
in detail.
\end{abstract}

\vspace{0.5cm}
\hspace{0.4cm}Keywords: A. thin films, D. charge density waves, D. optical
properties.

\vspace{0.5cm}

\section{Introduction}

The microwave, far infrared, and optical properties of charge density
wave (CDW) in quasi one--dimensional conductors (for reviews of CDW
physics see [1--3]) are characterized by the large polarization which
arises when the CDW is deformed along the chain direction.

At low temperatures, the optical dielectric function for the chain
direction is essentially that of a polar crystal -- albeit with very
different frequency scales due to the heavy Fr\"ohlich mass [4,5].
Together with the extreme anisotropy this leads to peculiar
far infrared responses.

The elementary excitations of
CDW which couple directly to the electric field are the phasons. Their
theoretical discussion has a long history [5--15]. Phasons give
rise to phason--polaritons [15] when coupled to the electromagnetic
field.

In this paper we study theoretically surface phason--polaritons in
CDW films which can now be fabricated [16].

Surface polaritons are well known in conjunction with optical phonons,
plasmons and magnons and have been reviewed in [17--20].
The present paper is directly related to the work of Kliewer and Fuchs
[21] on surface phonon--polaritons in isotropic films. We generalize
their work to anisotropic surface phason--polaritons involving phason
dispersion functions calculated in [15].
Uniaxial half--space polariton problems have been considered in
[22--24] (cf. also [18]) for other polariton mechanisms and mainly for
non--retarded interaction.

We will also investigate the effects of finite temperatures when
thermally excited quasi--particles (qp) cause surface phason--polariton
damping and level shifts.

It is excepted that surface phason--polaritons can be investigated
experimentally by attenuated total reflection (ATR) [25], possibly
also by electron energy loss spectroscopy (EELS) [26], and by low energy
electron diffraction (LEED) [27]. The guided waves which we find along with
the surface phason--polaritons can be measured in microwave cavity experiments
similar to those on superconducting films [28].

\section{Basic Theory}

The bulk results for the dielectric tensor $ \epsilon_{ij} ({\bf
q},\omega )$ of a CDW given as equations (41--44) in [15] refer to
long wave
lengths $ q \ll \xi ^{-1}_0 = \Delta /v_F $ ($\Delta$: half gap, $v_F$:
Fermi velocity) when $ {\bf q}$ points
into chain direction. The amplitude coherence length $ \xi _0 $ of
CDW is rather small away from $ T_c $. Typical values are $ \xi _0
\approx 3  nm $ due to the large value of the half gap $ \Delta  $.
The bulk dielectric functions can thus be used for wave lengths $
q < 10^6$ $cm^{-1}$ and for film thicknesses $ 2d > 30 nm$. The geometry
under study is shown in Fig. 1.

\begin{figure}[ht]
\begin{center}
\epsfig{file=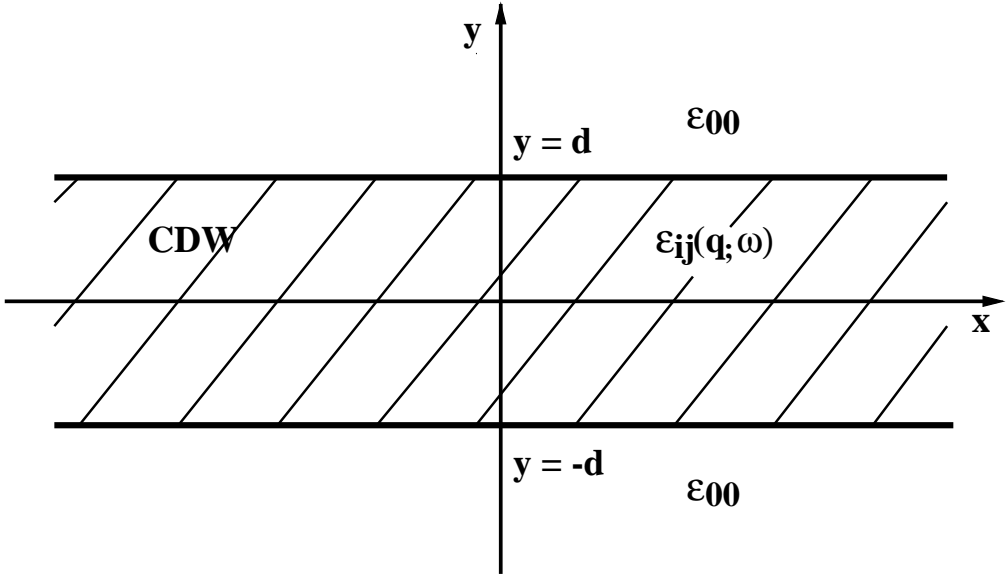,width=380pt}
\caption{\it{ CDW film of thickness $2d$ and with dielectric function
	$\epsilon_{ij}({\bf q},\omega)$ embedded in a medium of dielectric
	constant $\epsilon_{00}$. The conducting CDW--axis is chosen as
	$x$--axis.}}
\end{center}
\end{figure}

The film is embedded in a dielectric medium with dielectric constant $
\epsilon _{00}.$ The chain direction of the CDW system lies in the film
and is chosen as
the x--axis. The corresponding wave number is henceforth called
$ q$. For $ d \rightarrow \infty $ , the limiting case of a half
space is obtained. We consider non--radiative electric modes
(P--polarization). The electric field inside the film then has
components $E_x $ and $E_y $ and the magnetic field is transverse,
lying perpendicular to the conducting axis $ (B_z \not= 0) $. The
three--dimensional wave vector then consists of the longitudinal
component $ q_x = q $ and the transverse component $ q_y \equiv k
= - i \alpha$. As shown in [15], the generalization of the
solvability relation in [21] to the present anisotropic case is

\begin{equation}
\frac{\omega ^2}{c^2}\left (\epsilon _{xx} \epsilon _{yy} - \epsilon
_{xy}^2 \right ) + \alpha ^2 \epsilon _{yy} - q^2 \epsilon _{xx} +
2iq \alpha  \epsilon _{xy} = 0.
\end{equation}

\noindent
The matching condition for the electromagnetic fields takes on the
form

\begin{equation}
\frac{\epsilon _{00}}{\alpha _0} = - \left\{ \begin{array}{c}
\coth  (\alpha d)\\
\tanh  (\alpha d) \end{array} \right\}
\quad
\left[ \frac{\epsilon_{xx} \epsilon_{yy} - \epsilon_{xy}^2}{\alpha
 \epsilon_{yy} + iq \epsilon_{xy}} \right ].
\end{equation}

\noindent
The coth--equation gives the high frequency and the tanh--equation the
the low frequency branches.
In comparison to [15], we changed the notation of the dielectric
functions: $ \epsilon _{ll} \rightarrow \epsilon _{xx}, \epsilon_{t1t1}
\rightarrow \epsilon _{yy}, \epsilon _{lt} \rightarrow \epsilon
_{xy}.$
For surface polaritons, $ \alpha $ is positive and defines the decay
length in y--direction. The oscillating guided wave modes have imaginary
$ \alpha  = ik $. The light line of the embedding medium is $ \alpha
_0 = 0 $ where $ \alpha _0 $ is defined by

\begin{equation}
\alpha ^2_0 = q^2 - \frac{\omega ^2}{c^2} \epsilon_{00}.
\end{equation}

\noindent
Eliminating $ \epsilon _{yy} $ from (2), using (1) gives the simpler
formula

\begin{equation}
\frac{\epsilon _{00}}{\alpha _0} = - \left\{ \begin{array}{c} \coth
(\alpha d)\\
\tanh (\alpha  d) \end{array} \right\}
\quad
\left[ \frac{q\epsilon _{xx}- i\alpha \epsilon _{xy}}{i\omega
^2\epsilon_{xy}/c^2+ q\alpha} \right]
\end{equation}

\noindent
for the matching condition.

Equations (1), (3), and (4) determine the desired dispersion relation $
\omega  = \omega (q) $ by eliminating $ \alpha $. This is clearly
a complicated task. Enormous simplifications occur in the limit $ T
\rightarrow 0 $ when all $ qp $ effects freeze out in the CDW
dynamics.

\section{CDW Dielectric Tensor at Zero Temperature}

The relevant components of the dielectric tensor at zero temperature
according to [15] have the simple form

\begin{eqnarray}
\epsilon _{xx}(q,k;\omega ) = \epsilon _0 + \epsilon _\Delta \eta_1
(q) - \frac{c^2_l \kappa^2\eta(q)}{\Omega^2(q,k;\omega )},\\ [3mm]\nonumber
\epsilon _{yy } (q,k;\omega ) = \epsilon _t ,\,\,\,\ \epsilon _{xy} =
\epsilon _{yx} = 0,\\[3mm] \nonumber
\Omega ^2 \equiv \omega ^2 - \omega ^2_0 - c^2_l q^2 \eta(q) +
c^2_t\alpha ^2.
\end{eqnarray}

\noindent
Here,

\begin{equation}
\epsilon _\Delta  = \frac{v^2_F\kappa ^2}{6\Delta ^2_0} \gg \epsilon
_0
\end{equation}

\noindent
is the dielectric constant from virtual transitions across the gap
$ 2\Delta_0 $ [5].

\noindent
In (6), $\kappa$ is the Thomas--Fermi wave number
and $ 2\Delta _0 $ the zero temperature gap. $ \epsilon _\Delta $
is much larger than the lattice dielectric constant $ \epsilon _0
$ in chain direction. Equation (6) is a valid form for $ \omega <
\Delta _0 $. $\eta$ and $ \eta_1 $ are dispersive corrections which become
unity for $ q \rightarrow 0 $ [15]. The phason velocity $ c_l $ at
zero temperature and for the chain direction is given in terms of
Fermi velocity $ v_F $, Peierls phonon frequency $ \omega _Q $,
electron phonon coupling constant $ \lambda  $, and the half gap $
\Delta_0 $ as [5] :

\begin{equation}
c^2_l = \frac{\lambda \omega ^2_0}{4\Delta ^2_0} v^2_F \ll v^2_F.
\end{equation}

\noindent
The characteristic frequencies of the zero temperature theory are
the frequency $\omega_{LO} $ of the longitudinal optical (LO) phason

\begin{equation}
\omega ^2_{LO} = \frac{c^2_l \kappa ^2}{\epsilon _\Delta } \equiv
\frac{3}{2}\lambda \omega ^2_Q,
\end{equation}

\noindent
and the pinning frequency $ \omega _0 $ which usually satisfies

\begin{equation}
\omega _0 \ll \omega_{LO}.
\end{equation}

\noindent
This is consistent with the clean limit assumption underlying [15].
The frequency $\omega_{LO}$ is of order $10^{13} s^{-1}$ and thus
much smaller than usual optical phonon frequencies. For this reason,
the transverse lattice dielectric function $\epsilon _t$ will be taken
as constant in the considered frequency range. This is a strong
assumption in view of the possible appearance of bound collective
mode resonances in the FIR [29].

A useful dimensionless measure of the wave number $q$ for
phason--polaritons then is (c: light velocity)

\begin{equation}
Z = \frac{c^2q^2}{\omega_{LO}^2 \epsilon _t}.
\end{equation}

\noindent
Z is of order unity for the present purposes, i.e., $q \le 10^3$
$cm^{-1}$. From $c^2_l q^2 \equiv \epsilon _t\omega ^2_{LO} Z (c_l/c)^2
$ it is seen that all dispersive effects inside the CDW dielectric
functions are irrelevant. Thus one obtains at zero temperature

\begin{equation}
\epsilon _{xx}(\omega ) = \epsilon _\Delta \left (1 - \frac{\omega
_{LO}^2}{\omega ^2 - \omega _0^2} \right) \cong \epsilon _\Delta
\left (\frac{\omega ^2 - \omega_{LO}^2}{\omega ^2 - \omega _0^2}
\right ),
\end{equation}

\noindent
Equation (11) is identical with the dielectric function of a polar crystal,
$ \omega _0$ playing the role of the transverse optical frequency
$\omega_{TO}$ and $\epsilon _\Delta $ that of the dielectric constant
$\epsilon_\infty$. However, (11) refers only to the chain direction.
The Lyddane--Sachs--Teller relation

\begin{equation}
\epsilon ^* = \epsilon _\Delta	\quad  \frac{\omega^2 _{LO}}{\omega_0^2},
\end{equation}

\noindent
does not give the static dielectric constant of CDW --which
does not exist-- but the "plateau" dielectric constant $ \epsilon
^*$ [30] as explained in [31]. The decisive difference to the
usual phonon polariton case is the strong anisotropy $ \epsilon
_{xx} \not= \epsilon _t$ and the different frequency scale.
With these simplification the solvability relation (1) reduces to

\begin{equation}
\alpha ^2 = \frac{\epsilon _{xx}}{\epsilon _t} \left(q^2 -
\frac{\omega ^2}{c^2}\epsilon _t \right).
\end{equation}

\noindent
The matching condition becomes

\begin{equation}
\frac{\epsilon _{00}}{\alpha _0} = - \left\{ \begin{array}{c} \coth
(\alpha  d)\\
\tanh (\alpha  d) \end{array} \right\}
\quad
\frac{\epsilon _{xx}}{\alpha }.
\end{equation}

\noindent
Equations (3), (13), and (14) completely determine the excitation
spectrum.

\section{Spectra at Zero Temperature}

We will display the phason--polariton spectra in the $ W-Z $ plane. $ Z $
is given by (10) and the reduced squared frequency $ W $ is defined
by

\begin{equation}
W = \frac{\omega ^2}{\omega_{LO}^2}.
\end{equation}

\noindent
The existence of surface phason--polaritons requires

\begin{equation}
\epsilon _t > \epsilon _{00},
\end{equation}

\noindent
which can be fulfilled. The $ W-Z $ quadrant is devided into several
regions by the light lines $ W = \epsilon_t Z/\epsilon _{00}$ ($\alpha
^2_0 = 0$) and $ W = Z$ ($\alpha ^2 = 0$) plus horizontal lines $ W
= 1 $ and $ W = W_0 $ where $ \alpha ^2 $ is also zero due to
vanishing of $ \epsilon_{xx} $. The quantity $ W_0 $ means $\omega^2_0
/\omega_{LO}^2 $. Fig. 2 shows this partitioning of
the $ W-Z $ plane for $ \epsilon _t = 2\epsilon _{00}$. The shaded
areas have $ \alpha ^2 > 0 $. They --potentially-- could contain
surface phason--polaritons. We call them region 1, 2, and 3, respectively.
The region to the left of the light line $ \alpha _0 = 0 $ belongs
to $ \alpha ^2_0 < 0 $ , i.e., to radiative solutions which we do
not consider. The remaining regions I, II, and III have $ \alpha ^2
< 0. $ They --possibly-- contain guided wave solutions. The answers
to these questions require detailed investigations of (14), paying
special attention to the sign of $ \alpha  $ and of $ \epsilon _{xx}$.

\begin{figure}[ht]
\begin{center}
\epsfig{file=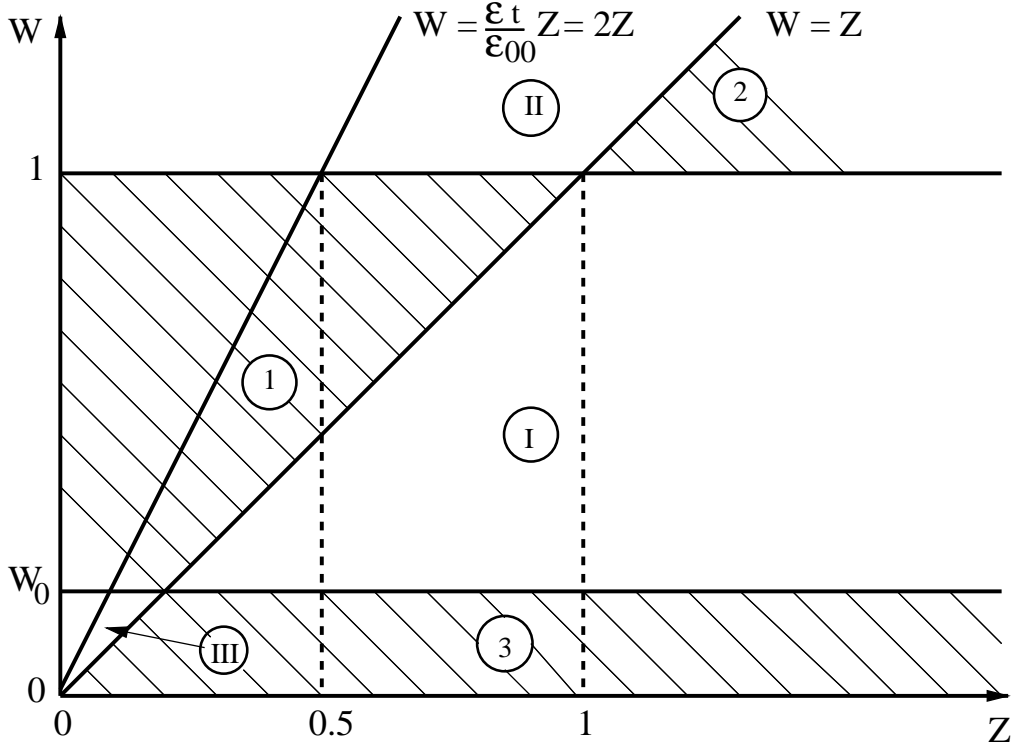,width=380pt}
\caption{\it{ Partitioning of reduced frequency versus wave number plane
	for the polaritonic response of a CDW film at zero temperature.
	$W=\omega^2/\omega_{LO}^2$ measures squared frequency and $Z=c^2q^2
	/(\omega_{LO}^2 \epsilon_t)$ squared wave number in chain direction.
	$W=Z$ and $W=2Z$ are the light lines for CDW film and embedding
	medium, respectively. Regions 1, 2, and 3 sustain surface
	phason--polaritons ($\alpha^2>0$) while regions I, II, and III allow
	for guided waves ($\alpha^2<0$). The region to the left of $W=2Z$
	contains the radiative solutions. $W_0$ measures CDW--pinning.}}
\end{center}
\end{figure}

We introduce dimensionless parameters according to

\begin{equation}
h = \frac{\epsilon _t}{\epsilon_{00}},\,\,\,\, g_0 = \frac{\epsilon _\Delta
}{\epsilon _{00}},\,\,\,\, D_0=\sqrt{\epsilon _\Delta }d \omega_{LO}/c.
\end{equation}

\noindent
For a film of thickness 2d = 1000 nm [16], a typical value of $ D_0
$ would be 0.2.

\subsection{Surface Phason--Polaritons}

Surface phason--polaritons exist only in region 1 where (14) translates into

\begin{equation}
\sqrt{\frac{(W-W_0)(W-Z)}{g_0
(hZ-W)(1-W)}} = \left\{ \begin{array}{c}
\coth X\\
\tanh X,  \end{array} \right.
X \equiv D_0 \sqrt{\frac{(1 -W)(W- Z)}{W - W_0}}.
\end{equation}

\noindent
For each thickness $D_0$, there is one surface phason--polariton branch
from the coth--equation and one from the tanh--equation separated
by the limiting curve for $ D_0 \rightarrow \infty $ which corresponds
to the half space problem. The coth--branch ends at the line $ W = 1
$
and the tanh--branch at the line $ W = Z $ , as shown in Fig. 3. Both
branches start at $ (Z = W_0/h, W = W_0) $ and have $ W = hZ $ as
tangent there.
The limiting curve for $ D_0 \rightarrow \infty $ is given implicitly
by

\begin{equation}
\left(1 - \frac{1}{g_0} \right) W^2 - \left(1 + hZ - \frac{Z +
W_0}{g_0}\right) W =
\left(\frac{W_0}{g} - h \right)Z.
\end{equation}

\noindent
The quantity $ g_0 $ can be much larger than unity. Then the
coth--branches (one for each $ D_0 $ ) are squeezed to the $ \alpha _0
= 0 $ line and thus vanish.

There are no solutions in regions 2 and 3.

\begin{figure}[ht]
\begin{center}
\epsfig{file=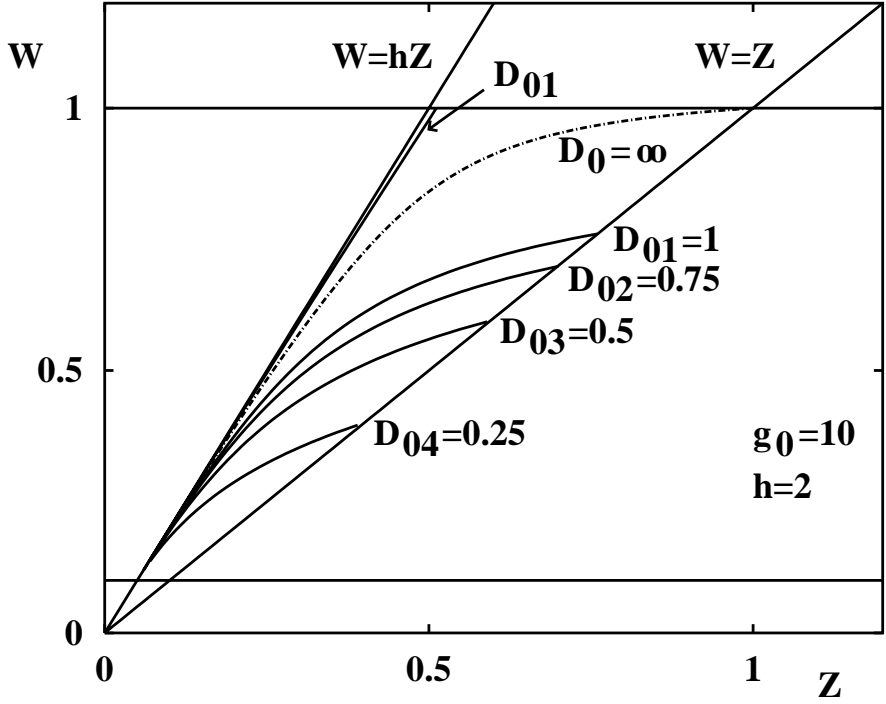,width=380pt}
\caption{\it{Surface phason--polariton branches of a CDW film for several
	values of the scaled thickness $D_0$. The limiting curve $D_0
	\rightarrow\infty$ (dash--dotted line) separates the region of even
	solutions (to its right) from the region of odd solutions.
	The quantities $g_0$ and $h$ are the high frequency CDW dielectric
	constants along and transverse to the chains in relation to the
	embedding medium dielectric constant, respectively.}}
\end{center}
\end{figure}

\subsection{Guided Waves}

In the regions I, II, and III, $ \alpha ^2 $ is negative, i.e., the
transverse wave number $ k $  is real and the modes oscillate across
the film. For imaginary $ \alpha $, we find in region I a series
of solutions ( p= 0,1,2,3,...)

\begin{equation}
D_0 \sqrt{\frac{(Z - W)(1 - W)}{W - W_0}} = p\frac{\pi}{2} + \arctan
\sqrt{\frac{(Z - W)(W - W_0)}{g_0(hZ - W)(1 - W)}}.
\end{equation}

\noindent
Modes with p = 2m (m= 0,1,2,...) are from the former tanh--equation and
those with p = 2m + 1 are from the coth--equation. The $p=0$ mode is
the continuation of the corresponding surface phason--polariton.
The $p > 0$ modes start linearly at $(Z = W_0/h, W = W_0)$. All modes go
asymptotically towards $\omega_{LO}$ for large $q$. Fig. 4 shows
these guided wave modes.
The slopes of the $ p > 0$ modes near $Z = W_0/h$ are given by
$D_0^2 (1-W_0)/ (D_0^2(1-W_0) + p^2 (\pi /2)^2)$ and decrease
with increasing order p.

\begin{figure}[ht]
\begin{center}
\epsfig{file=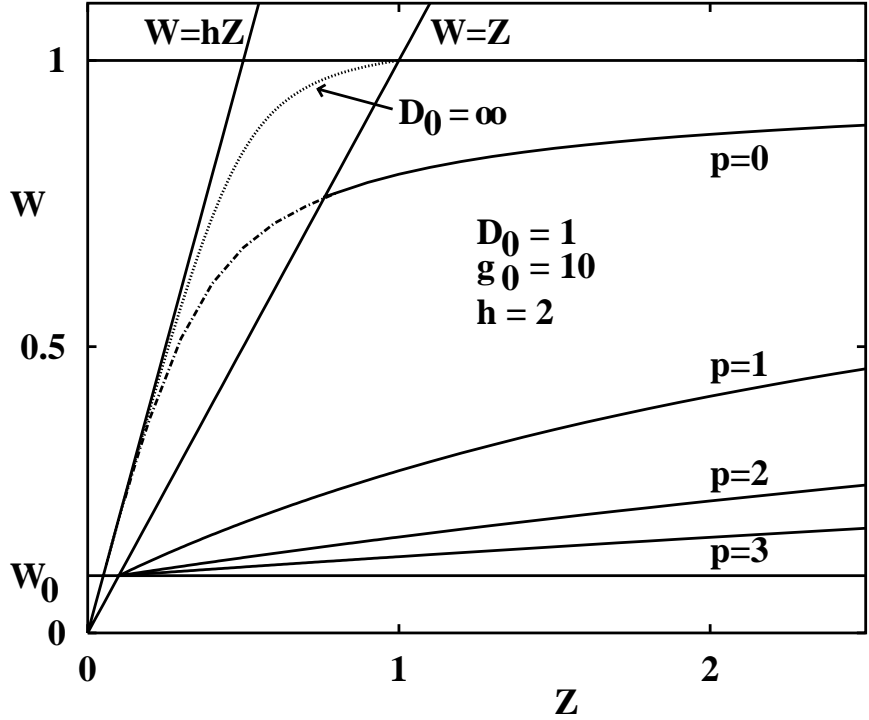,width=380pt}
\caption{\it{Guided wave modes of region I in a CDW film for a fixed (scaled)
	thickness $D_0=1$ and for several mode numbers $p$. The $p=0$
	mode is the continuation of the surface phason--polariton in
	region 1 (dash--dotted line). The quantities $g_0$ and $h$ are the high
	frequency CDW dielectric constants along and transverse to the chains
	in relation to the embedding medium dielectric constant, respectively.}}
\end{center}
\end{figure}

Guided waves also exist in region II but not in III. Clearly, there
must be a continuation of the coth--surface--polariton branch of
region 1 into region II. This is the $n=1$ solution of

\begin{equation}
-D_0 \sqrt{\frac{(W-Z)(W-1)}{W - W_0}} = - n \frac{\pi }{2} + \arctan
\sqrt{\frac{(W-Z)(W-W_0)}{g_0(hZ-W)(W-1)}}.
\end{equation}

\noindent
The modes with $n=2m \,(m=1,2,3,\ldots)$ are from the former
tanh--equation, the others ($n=2m + 1, m = 0,1,2,3, \ldots$) are
from the coth--equation. All modes with  $n \ge 2$ start at discrete
points $(Z_m, W_m = h Z _m)$ on the $\alpha _0 = 0 $ line and go
asymptotically to the $W=Z$ line. The points $Z_m$ solve the equation

\begin{equation}
(h-1) Z_m (hZ_m-1) = n^2 \left( \frac{\pi }{2D_0}\right)^2  (hZ_m
- W_0).
\end{equation}

\noindent
These modes, therefore, start at rather high frequencies, the lowest
one at $Z_2 \cong 3$ for our choice $D_0 = 1$ and $h = 2$.

Fig. 5 summarizes all our results for the excitation spectra in CDW
films at zero temperature.

\begin{figure}[ht]
\begin{center}
\epsfig{file=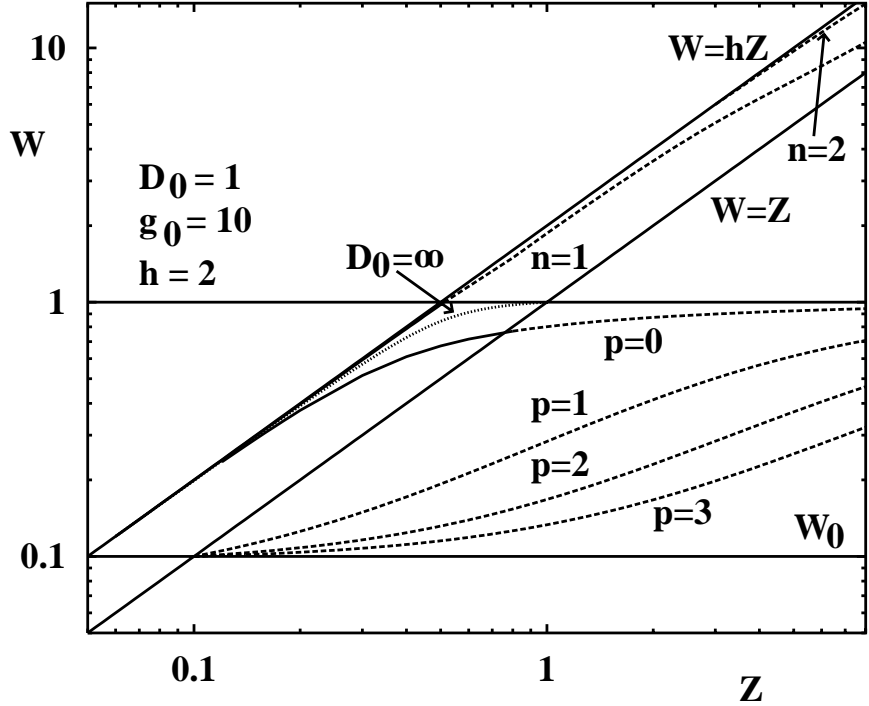,width=380pt}
\caption{\it{Survey of all polaritonic excitations of a CDW film: Surface
	phason polaritons in region 1 (full line) and guided wave modes in
	regions I and II (broken lines, numbered by integers $p$)
	for a fixed (scaled) thickness $D_0=1$.
	The quantities $g_0$ and $h$ are the high frequency CDW dielectric
	constants along ($\epsilon_\Delta$) and transverse to
	the chains ($\epsilon_t$) in relation to the
	embedding medium dielectric constant $\epsilon_{00}$, respectively.
	The limiting curve for $D_0 \rightarrow\infty$ (dotted line)
	is the surface phason--polariton of a CDW half space.}}
\end{center}
\end{figure}

\vspace{0.8cm}

\subsection{Neglect of Retardation}

It is interesting to consider the limit $ c \rightarrow \infty$, i.e.,
when retardation can be neglected. In this limit, only region I
survives leaving only guided wave modes. They follow from

\begin{equation}
\sqrt{\tilde{Z}\frac{1-W}{W-W_0}} = p\frac{\pi }{2} + \arctan
\sqrt{\frac{W-W_0}{g_0h(1-W)}}, \,\,\,\, p = 0,1,2,3,....
\end{equation}

\noindent
Here, $\tilde{Z}$ combines wave number q and film thickness 2d
according to

\begin{equation}
\tilde{Z} = \frac{\epsilon _\Delta }{\epsilon _t} (qd)^2.
\end{equation}

\noindent
The p = 0 curve for small $\tilde{Z}$ was calculated in [15] and has
the characteristic behaviour $\omega  \sim \sqrt{q}$ of a
two--dimensional plasmon for $\omega _0 = 0$. The modes with $ p \ge
1$ are low lying and are acoustic for $\omega_0 = 0$ as shown in Fig. 6.

\begin{figure}[ht]
\begin{center}
\epsfig{file=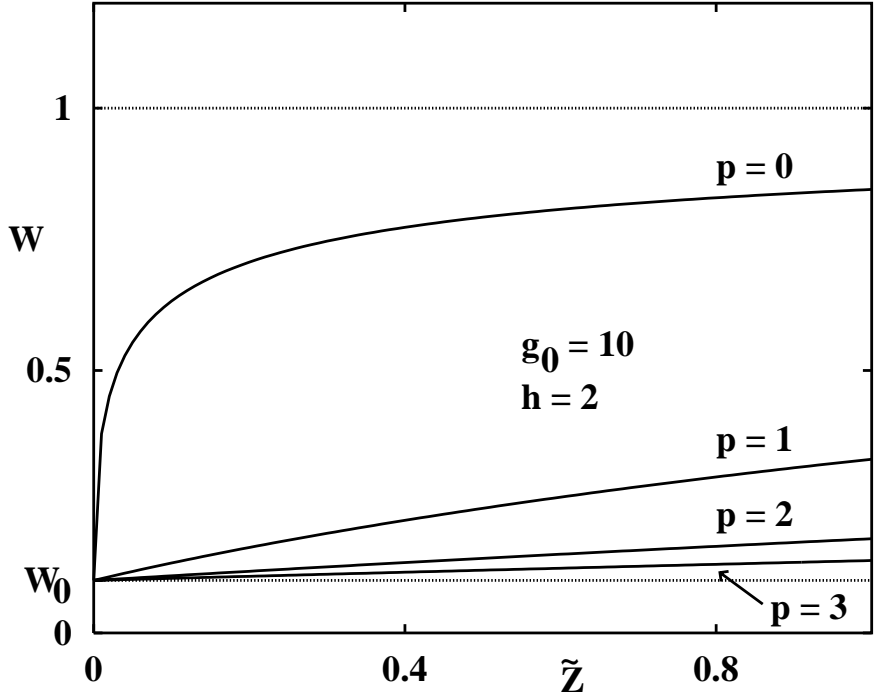,width=380pt}
\caption{\it{Non--retarded limit of polaritonic response of CDW films at zero
	temperature. The wave number scale is $\tilde{Z}=\epsilon_\Delta
	q^2d^2/\epsilon_t$ ($2d$: film thickness).
	The quantities $g_0$ and $h$ are the high frequency CDW dielectric
	constants along ($\epsilon_\Delta$) and transverse to
	the chains ($\epsilon_t$) in relation to the
	embedding medium dielectric constant $\epsilon_{00}$, respectively.
	The limiting curve for $D_0 \rightarrow\infty$ (dotted line)
	is the surface phason--polariton of a CDW half space.}}
\end{center}
\end{figure}

\section{Spectra at Finite Temperatures}

At finite temperatures, the expressions for the components of the
dielectric tensor become very complicated and contain additional
frequency and wave number dependencies. They also depend on elastic and
inelastic scattering rates of the $qp$.

At sufficiently high temperatures of about $T_c/2$ or higher, the
inelastic scattering becomes relevant and the corresponding formulae
for $\epsilon_{ij}$ are given as eq. (62) in [15]. Inserting these
expressions into the solvability condition (1), a surprising number
of cancellations occur. A further significant simplification is the
neglect of dispersion in CDW functions by utilizing the argument
of Sec. 3. In the end, this simplification amounts to the following
approximations for the dielectric tensor

\begin{eqnarray}
\epsilon _{xx} \rightarrow \epsilon _\Delta  (T) \left[ 1 +
\frac{i\omega _l(T)}{\omega } - \frac{\omega _{LO}^2(T)}{\Omega^2(T)}
\right],\\ \nonumber
\epsilon _{yy} \rightarrow \epsilon _t \left[ 1 + \frac{i\omega
_t(T)}{\omega } \right],\,\,\,\, \epsilon _{xy} \rightarrow 0.
\end{eqnarray}

\noindent
Here, $\omega _{l,t} (T)$ are the dielectric relaxation frequencies
of $qp$ for longitudinal and transverse directions, respectively:

\begin{equation}
\omega _l (T) = \frac{4\pi  \sigma^{(0)}_l(T)}{\epsilon _\Delta (T
)},\,\,\,\, \omega _t(T) = \frac{4\pi \sigma ^{(0)}_t(T)}{\epsilon
_t}.
\end{equation}

\noindent
The dc--conductivities $\sigma ^{(0)}(T)$ are activated and account
for the main temperature dependence. They are
explicitly given by equation (57) in [15]. We will further assume a
quasi--one--dimensional situation when $\omega _t \ll \omega _l$
holds. Then the solvability relation (1) reduces to

\begin{equation}
\alpha ^2  = \frac{\epsilon _\Delta (T)}{\epsilon _t} \left[ 1 +
\frac{i\omega _l(T)}{\omega } - \frac{\omega _{LO}^2 (T)}{\Omega ^2} \right]
\left\{ q^2 - \frac{\omega ^2}{c^2} \epsilon _t \right\}.
\end{equation}

\noindent
This expression must be inserted into (14) to get the spectra $\omega
 = \omega (q,T)$. It is noted that $\omega$ becomes complex at finite
temperatures and that the imaginary part of $\omega$ measures polariton
damping.

Temperature dependencies appear not only in $\omega _l$ but also in
$\omega_{LO}, \Omega ^2$, and  $\epsilon _\Delta$.
The temperature dependent $ LO $ phason frequency is given in terms
of its zero temperature value $\omega_{LO}$ and the backflow parameter
$b_0$ as

\begin{equation}
\omega _{LO}^2 (T) \equiv \frac{c_l^2(T)\kappa ^2}{\epsilon _\Delta (T)}
(1-b_0)^2
\equiv \left\{ \frac{\epsilon _\Delta }{\epsilon _\Delta (T)}
\left( \frac{\Delta _0}{\Delta (T)} \right)^2 (1-b_0)^2\right\}
\omega _{LO}^2.
\end{equation}

\noindent
The expression for $\Omega^2$ is

\begin{equation}
\Omega^2(T) = \omega ^2 + i\omega \gamma_0(T) - \omega _0^2.
\end{equation}

\noindent
The phason damping $\gamma_0$ from $qp$ backscattering is defined by
equation
(60) in [15]. We estimate its value by introducing a cut--off $\nu
_c \ll k_BT $ near the band gap, assuming $\Delta  >k_BT$ and $\nu
_b = \nu $ for the $qp$ scattering rates (cf. [15]). This gives

\begin{equation}
\gamma_0(T) \approx \frac{\nu }{4\Delta } \, \frac{\lambda \omega
_Q^2}{k_BT}\, \left| \ln \left( \frac{\nu_c}{k_BT}\right) \right|
\exp(-\frac{\Delta }{k_BT}).
\end{equation}

\noindent
Thus $\Gamma  \equiv \gamma_0 / \omega_{LO}
\equiv c_l\nu|\ln(\nu_c/k_BT)|\exp(-\Delta/k_BT)/(v_Fk_BT\sqrt{6})\ll 1$
holds and direct phason damping is not important
in the frequency and temperature range
under study. The phason--polariton damping stems from $qp$ dissipation
expressed by $ \omega _l. $

\noindent
The temperature dependence of $ \epsilon _\Delta (T) $
cannot be neglected even when $ \Delta = \Delta _0 $ is considered.
It is contained (cf. [15]) in

\begin{equation}
\frac{\epsilon _\Delta (T)}{\epsilon _\Delta} = \left( \frac{\Delta
_0}{\Delta (T)} \right) ^2 \frac{3\pi }{2 \sqrt{z}} \sum^\infty_{n
= 0} \frac{1}{[(2n + 1)^2 \pi ^2/(4z) + 1]^{5/2}} \equiv \left(
\frac{\Delta _0}{\Delta (T)} \right)^2 g_1(T).
\end{equation}

\noindent
The quantity $ z $ is defined by

\begin{equation}
z = \left( \frac{\Delta }{2k_BT} \right)^2.
\end{equation}

\noindent
For a qualitative study of finite temperature effects we adopt a
simple model in which $\Delta (T) $ is fixed at $ \Delta _0 $. This
is not unrealistic for CDW up to $20 \% $ $qp$ fraction ($N = 0.8$).
As in [15] we then find

\begin{equation}
l(T)  \equiv \frac{\omega _l(T)}{\omega _{LO}} = (1 - N)\, 8 \,\sqrt
{\pi } \left( \frac{k_BT}{\Delta _0} \right)^{3/2}
\frac{\sigma _N}{\epsilon _\Delta (0)\omega_{LO}g_1} \equiv
\left( \frac{T}{T_0} \right)^{3/2} r_0(1 - N)/g_1.
\end{equation}

\noindent
In (33), $\sigma _N $ denotes the normal state conductivity. For a
reference temperature of $ T_0 = \Delta _0/k_B = 1000 K $, the
parameter $r_0 $ is estimated to be of order 10, the value we use  in the
figures. The
expansion of the $qp$ fraction $ 1-N $ for $ \Delta \gg k_BT $ is

\begin{equation}
1-N = \sqrt{4\pi \sqrt{z}} \exp(-2 \sqrt{z}),
\end{equation}

\noindent
and for $ \nu = \nu _b $, the backflow parameter $ b_0 $ is approximated
by

\begin{equation}
b_0 = \frac{1}{2 \sqrt{z}} (1-N).
\end{equation}

\subsection{Guided Waves at Finite Temperatures}

Using the same scaling as for $ T=0 $, the matching condition (14)
can be written for region I ($p=0,1,2,3,\ldots)$ as

\begin{eqnarray}
D(T) \sqrt{(Z-W) \left[ \frac{a(T)-W}{W-W_0}-\frac{il(T)}{ \sqrt{W}}
\right] } =
p\frac{\pi }{2}\\[3mm]\nonumber
 + \arctan
\sqrt{\frac{Z-W}{g(T)(hZ-W)\left[(a(T)-W)/(W-W_0)-il(T)/
\sqrt{W} \right] }}.
\end{eqnarray}

\noindent
Here, the principal values of the square roots and the $\arctan$
must be taken.
The continuation of the $ p=0 $ solution into region 1 gives the
surface phason--polariton from the former tanh--equation.
The following parameters which extend zero temperature
quantities are used in (36):

\begin{eqnarray}
D(T) & = & g_1(T)\,D_0,\\ \nonumber
g(T) & = & g_1(T)\,g_0,\\ \nonumber
a(T) & = & (1-b_0(T))^2/g_1(T).
\end{eqnarray}

\noindent
The quantity $a$ reduces to unity for $ T \rightarrow 0 $.
It is a measure of the change of $\omega_{LO}(T)$ with temperature. For
small $qp$ concentrations, the decrease of $g_1(T)$ with increasing
temperature makes
$\omega_{LO}(T)$ larger than the zero temperature value $\omega_{LO}$.
At even higher temperatures, the backflow parameter $b_0$ but also the
increased damping $l(T)$ and the decreased order parameter $\Delta(T)$
become relevant for the effective value of $\omega_{LO}(T)$: The LO--phason
softens and becomes overdamped. We do not display this region because
it is not properly covered by our temperature model.

The non--retarded version of this equation is

\begin{eqnarray}
\sqrt{\tilde{Z} g_1 \left[\frac{a-W}{W-W_0}-i\frac{l}{ \sqrt{W}} \right]
} = p\frac{\pi }{2}+ \arctan \sqrt{\left(hg
\left[\frac{a-W}{W-W_0}-i\frac{l}{ \sqrt{W}} \right]\right)^{-1} }.
\end{eqnarray}

\noindent
Corresponding results are shown in Fig. 7. It is seen that the fundamental
mode ($p=0$) is shifted to higher frequencies when the $qp$ fraction
increases. This reflects the increase of the LO--phason frequency which
has been described above. The guided waves remain underdamped but the
imaginary parts of its frequencies do become significant for $N \ge 0.9$. The
higher order modes ($p>0$) are much less affected by $qp$.

\begin{figure}[ht]
\begin{center}
\epsfig{file=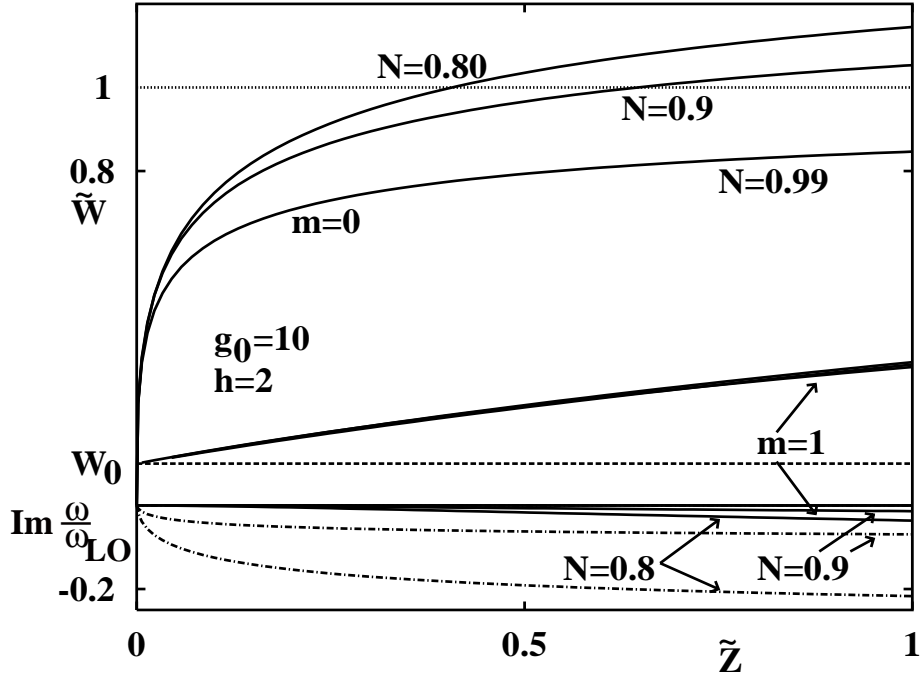,width=380pt}
\caption{\it{Non--retarded limit of polaritonic response of CDW films at
	finite temperatures. The wave number scale is $\tilde{Z}=
	\epsilon_\Delta q^2d^2/\epsilon_t$ ($2d$: film thickness)
	and $\tilde{W}$ means $(Re\omega/\omega_{LO})^2$.
	N is the condensate density and guided wave damping is displayed
	as $Im \omega/\omega_{LO} <0$.
	The quantities $g_0$ and $h$ are the high frequency CDW dielectric
	constants along ($\epsilon_\Delta$) and transverse to
	the chains ($\epsilon_t$) in relation to the
	embedding medium dielectric constant $\epsilon_{00}$, respectively.}}
\end{center}
\end{figure}

We like to point out that the surface and bulk phasons for $c \rightarrow
\infty$ studied in Section 7.2.3  of [15] appear in a different range
than the present phason--polaritons, namely at much smaller frequencies, larger
wave numbers, and higher temperatures ($\omega_l > \omega_{LO}$). Neglect
of internal CDW--dispersion is then not permitted.

\subsection{Surface Phason--Polaritons at Finite Temperatures}

Finally, we study the surface phason--polaritons in region 1 starting from the
appropriate formulae for region 1:

\begin{eqnarray}
\sqrt{\frac{W-Z}{g(T)(hZ-W)\left[(a(T)-W)/(W-W_0)-il(T)/
\sqrt{W} \right] }} = \left\{ \begin{array}{c}
\coth X(T)\\
\tanh X(T),  \end{array} \right. \\[3mm]\nonumber
X(T) \equiv D(T) \sqrt{(W-Z) \left[ \frac{a(T)-W}{W-W_0}-\frac{il(T)}
{ \sqrt{W}}\right] }.
\end{eqnarray}

\noindent
Unfortunately, these transcendental equations are difficult to solve,
even on a computer.
Therefore, we limit ourselves to the half space case $D(T)\rightarrow
\infty$ when a simple fifth order polynomial in the frequency
$\omega \propto \sqrt{W}$ remains:

\begin{equation}
g(T)(hZ-W)\left[\sqrt{W}(a(T)-W)-il(T)(W-W_0) \right] =
\sqrt{W}(W-W_0)(W-Z).
\end{equation}

\begin{figure}[htb]
\begin{center}
\epsfig{file=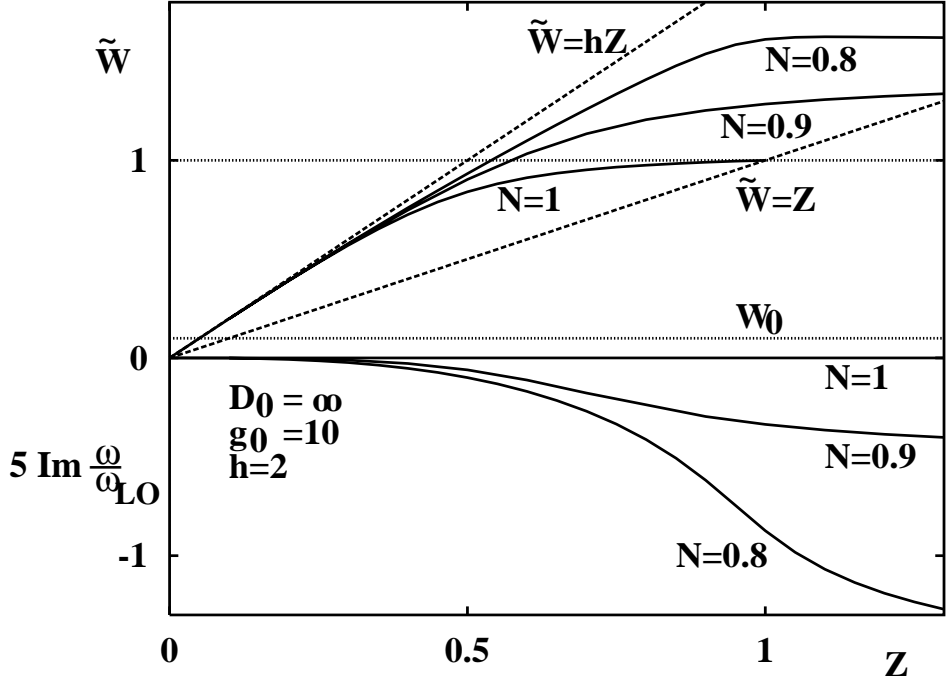,width=380pt}
\caption{\it{Surface phason--polaritons in a thick CDW film at finite temperatures. 
	$\tilde{W}$ means $(Re\omega/\omega_{LO})^2$. 
	N is the condensate density and polariton damping is displayed 
	as $5 Im \omega/\omega_{LO} <0$. 
	The quantities $g_0$ and $h$ are the high frequency CDW dielectric 
	constants along and transverse to the chains in relation to the 
	embedding medium dielectric constant, respectively.}}
\end{center}
\end{figure}

\noindent
Its surface phason--polariton solutions are shown in Fig. 8 for several $qp$
densities. In contrast to usual surface phonon--polaritons, the $qp$ cause a
significant temperature variation of the surface polariton frequencies.
The surface phason--polaritons are underdamped for the $qp$ densities
considered.

\section{Summary}
The strong anisotropy of the polar crystal like optical response of
quasi one--dimensional CDW has a distinguished influence on the coupled
excitations of the electromagnetic field and CDW phasons. These
excitations are grouped into surface phason--polaritons and guided
wave modes and their frequencies are considerably smaller than the
corresponding excitations involving optical phonons. This is due to
the large effective mass of CDW which leads to a low frequency of the
longitudinal optical phason. At finite temperatures, quasi--particles
cause damping and considerable level shifts of both surface phason
polaritons and guided wave modes. We have given a complete discussion
of these surface phason--polaritons and guided wave modes for the case
of P--polarization and for the conducting CDW axis lying inside the
film.

\section*{Acknowledgement}
The authors thank F. Gleisberg for many helpful discussions.

\newpage

{\large \bf References}

\begin{description}
\item [1]  Monceau P., in: {\it Electronic Properties of Inorganic
	   Quasi--One--Dimensional Materials}, (Edited by P. Monceau),
	   D. Reidel (1985)

\item [2] Gr\"uner G., Zettl A., {\it Phys. Rep.} {\bf 119}, 117 (1985)

\item [3] Gr\"uner G., {\it Rev. Mod. Phys.} {\bf 60}, 1129 (1988)

\item [4] Fr\"ohlich H., {\it Proc. Roy. Soc.} {\bf A223}, 296 (1954)

\item [5] Lee P.A., Rice T.M., Anderson P.W., {\it Solid State Commun.}
	 {\bf 14}, 703 (1974)

\item [6] Overhauser A.W., {\it Phys. Rev. B} {\bf 3}, 3173 (1971)

\item [7] Lee P.A., Fukuyama H., {\it Phys. Rev. B} {\bf 17}, 542 (1978)

\item [8] Kurihara Y., {\it J. Phys. Soc. Jpn.} {\bf 49}, 852 (1980)

\item [9] Artemenko S.N., Volkov A.F., {\it Zh. Eksp. Theo. Fiz.} {\bf 80}
	  2018 (1981), {\bf 81}, 1872 (1981) ({\it Sov. Phys. JETP}
	  {\bf 53}, 1050 (1981), {\bf 54}, 992 (1981))

\item[10] Nakane Y., Takada S., {\it J. Phys. Soc. Jpn.} {\bf 54}, 977 (1985)

\item[11] Wong K.Y.M., Takada S., {\it Phys. Rev. B} {\bf 36}, 5476 (1987)

\item[12] Artemenko S.N., Volkov A.F., {\it Synthetic Metals} {\bf 29},
	  F407 (1989)

\item[13] Virosztek A., Maki K., {\it Phys. Rev. B} {\bf 48}, 1363 (1993)

\item[14] Brazovskii S., {\it J. Phys. I France} {\bf 3}, 2417 (1993)

\item[15] Artemenko S.N., Wonneberger W., {\it J. Phys. I France} {\bf 6},
	  2079 (1996)

\item[16] van der Zant H.S.J., Mantel O.C., Dekker C., Mooij J.E.,
	  Traeholt C., {\it Appl. Phys. Lett.} {\bf 68}, 3823 (1996)

\item[17] {\it Polaritons} (Edited by Burstein E., de Martini F.),
	  Proc. Taormina Res. Conf., October 2--6, 1972, Pergamon Press (1973)

\item[18] Bryksin V.V., Mirlin D.N., Firsov Yu.A., {\it Usp. Fiz.
	  Nauk} {\bf 113}, 29 (1974) ({\it Sov. Phys. Usp.}
	  {\bf 17}, 305 (1974))

\item[19] Agranovich V.M., {\it Usp. Fiz. Nauk} {\bf 115}, 199 (1975)
	  ({\it Sov. Phys. Usp.} {\bf 18}, 99 (1975))

\item[20] {\it Surface Polaritons} (Edited by Agranovich V.M., Mills D.L.),
	  North--Holland Publ. Comp. (1982)

\item[21] Kliewer K.L., Fuchs Ronald, {\it Phys. Rev.} {\bf 144} 495 (1966)

\item[22] Agranovich V.M., Dubovskii O.A., {\it Fiz. Tverd. Tela} {\bf 7},
	  3054 (1966) ({\it Sov. Phys. Solid State} {\bf 7}, 2885 (1966))

\item[23] Dubovskii O.A., {\it Fiz. Tverd. Tela} {\bf 12}, 3054 (1970)
	  ({\it Sov. Phys. Solid State} {\bf 12}, 2471 (1970))

\item[24] Lyubimov V.N., Sannikov D.G., {\it Fiz. Tverd. Tela} {\bf 14},
	  675 (1972) ({\it Sov. Phys. Solid State} {\bf 14}, 575 (1972))

\item[25] Otto A., {\it Z. Phys.} {\bf 216}, 398 (1968)

\item[26] B\"orsch H., Geiger J., Stickel W., {\it Z. Phys.} {\bf 212},
	  130 (1968)

\item[27] Ibach H., {\it Phys. Rev. Lett.} {\bf 24}, 1416 (1970)

\item[28] Buisson O., Xavier P., Richard J., {\it Phys. Rev. Lett.}
	  {\bf 73}, 3153 (1994)

\item[29] Degiorgi L., Gr\"uner G., {\it Phys. Rev. B} {\bf 44}, 7820 (1991)

\item[30] Wonneberger W., {\it Synthetic Metals} {\bf 43}, 3793 (1991)

\item[31] Wonneberger W., in: {\it Physics and Chemistry of
	  Low--Dimensional Inorganic Conductors} (Edited by Schlenker C.,
	  Greenblatt M., Dumas J., van Smaalen S.), Nato Adv. Study
	  Inst., Les Houches, June 12--23, 1995, Plenum (1996)
\end{description}

\end{document}